\documentclass{article} \usepackage{spconf}
\ninept

\usepackage{graphicx}
\usepackage[cmex10]{amsmath}
\usepackage{amsfonts,amssymb,latexsym,bm}
\usepackage[T1]{fontenc} % handles underscore properly in bibliography
\usepackage[hidelinks]{hyperref}
\usepackage{cite}
\usepackage{enumerate,subcaption}
\usepackage[usenames]{color}
\usepackage{etoolbox} % supplies \newtoggle
\newtoggle{extra} % establish switch to include extra stuff
\togglefalse{extra} % uncomment to NOT include extra stuff

\usepackage{algorithm,algpseudocode}% http://ctan.org/pkg/{algorithms,algorithmx}

\algnewcommand{\Define}[1]{%
  \State \textbf{define:}
  \Statex \hspace*{\algorithmicindent}\parbox[t]{.8\linewidth}{\raggedright #1}
}
\algnewcommand{\Inputs}[1]{%
  \State \textbf{inputs:}
  \Statex \hspace*{\algorithmicindent}\parbox[t]{.8\linewidth}{\raggedright #1}
}
\algnewcommand{\Initialize}[1]{%
  \State \textbf{initialize:}
  \Statex \hspace*{\algorithmicindent}\parbox[t]{.8\linewidth}{\raggedright #1}
}

\newcommand{\textr}[1]{\textcolor{red}{#1}}

\renewcommand{\hat}{\widehat}

\renewcommand{\vec}[1]{\boldsymbol{#1}}

\newcommand{\hvec}[1]{\hat{\boldsymbol{#1}}}

\newcommand{\defn}{\triangleq}
\newcommand{\mat}[1]{\ensuremath{\begin{bmatrix}#1\end{bmatrix}}}

 \newcommand{\mc}[1]{\ensuremath{\mathcal{#1}}}
\newcommand{\Real}{{\mathbb{R}}}
\newcommand{\Complex}{{\mathbb{C}}}

\newcommand{\tran}{^{\text{\textsf{T}}}}
\newcommand{\herm}{^{\text{\textsf{H}}}}
\DeclareMathOperator{\E}{\mathbb{E}}

\DeclareMathOperator{\tr}{tr}

\DeclareMathOperator{\gdiag}{gdiag}
\DeclareMathOperator{\Diag}{Diag}

\DeclareMathOperator{\prox}{prox}
\DeclareMathOperator{\gprox}{gprox}
\renewcommand{\eqref}[1]{(\ref{eq:#1})}

\newcommand{\Figref}[1]{Figure~\ref{fig:#1}}
\newcommand{\figref}[1]{Fig.~\ref{fig:#1}}
\newcommand{\tabref}[1]{Table~\ref{tab:#1}}

\renewcommand{\algref}[1]{Alg.~\ref{alg:#1}}

\newcommand{\iter}{t}
\newcommand{\iters}{{t+1}}
\newcommand{\itero}{{t-1}}

\newcommand{\MMSE}{_{\text{\sf MMSE}}}

\newcommand{\true}{_0}

\let\OLDthebibliography\thebibliography
\renewcommand\thebibliography[1]{
  \OLDthebibliography{#1}
  \setlength{\parskip}{3pt}
  \setlength{\itemsep}{0pt plus 0.3ex}
}

\makeatletter
\def\endthebibliography{%
	\def\@noitemerr{\@latex@warning{Empty `thebibliography' environment}}%
	\endlist
}
\makeatother

\begin{document}
\setlength{\arraycolsep}{0.5mm}

\title{Expectation Consistent Plug-and-Play for MRI}
\name{Saurav K. Shastri$^\dagger$,
        Rizwan Ahmad$^*$, 
        Christopher A. Metzler$^\circ$, 
        and
        Philip Schniter$^\dagger$\thanks{This work was funded in part by the National Institutes of Health under grants R01HL135489 and R01EB029957, and by the National Science Foundation under grant CCF-1955587.}}
\address{$^\dagger$Dept. ECE, The Ohio State Univ., Columbus, OH, 43210, \{shastri.19,\,schniter.1\}@osu.edu\\
         $^*$Dept. BME, The Ohio State Univ., Columbus, OH, 43210, rizwan.ahmad@osumc.edu\\
         $^\circ$Dept. CS, The Univ. of Maryland, College Park, MD, 20742, metzler@umd.edu}

\maketitle

%%%%%%%%%%%%%%%%%%%%%%%%%%%%%%%%%%%%%%%%%%%%%%%%%%%%%%%%%%%%%%%%%%%%%%%%%%%%%%%%
\begin{abstract}
For image recovery problems, plug-and-play (PnP) methods have been developed that replace the proximal step in an optimization algorithm with a call to an application-specific denoiser, often implemented using a deep neural network.
Although such methods have been successful, they can be improved.
For example, the denoiser is often trained using white Gaussian noise, while PnP's denoiser input error is often far from white and Gaussian, with statistics that are difficult to predict from iteration to iteration.
PnP methods based on approximate message passing (AMP) are an exception, but only when the forward operator behaves like a large random matrix.
In this work, we design a PnP method using the expectation consistent (EC) approximation algorithm, a generalization of AMP, that offers predictable error statistics at each iteration, from which a deep-net denoiser can be effectively trained.
\end{abstract}

\section{Introduction}
Magnetic resonance imaging (MRI) is a medical imaging approach that uses magnetic fields to create detailed anatomical images.
Although MRI provides excellent soft-tissue contrast without the use of ionizing radiation, it takes a long time to fully sample the measurement space.
Thus, it is common to take relatively few measurements and apply sophisticated post-processing to reconstruct an accurate image.
Although our paper focuses on MRI, the methods we propose apply to any application where the goal is to recover a signal from undersampled Fourier measurements.

\begin{figure}[t]
\centering
\includegraphics[width=0.4\columnwidth]{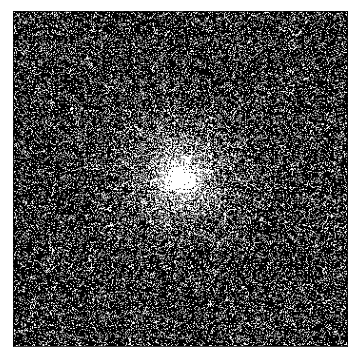}
\caption{A variable-density sampling mask at $R=4$.}
\label{fig:mri_mask}
\vspace{-5mm}
\end{figure}

The measurements $\vec{y}\in\Complex^{CM}$ collected in $C$-coil MRI, known as ``k-space'' measurements, can be modeled as
\begin{align}
\vec{y} 
&= \vec{Ax}\true + \vec{w} 
\text{~~with~~}
\vec{A}=\mat{\vec{MF}\Diag(\vec{s}_1)\\[-2mm]\vdots\\\vec{MF}\Diag(\vec{s}_C)}
\label{eq:y},
\end{align}
where 
$\vec{x}\true\in\Complex^N$ is a vectorized version of the $N$-pixel image we wish to recover,
$\vec{F}\in\Complex^{N\times N}$ is a unitary 2D discrete Fourier transform (DFT),
$\vec{M}\in\Real^{M\times N}$ is a sampling mask formed from $M$ rows of the identity matrix $\vec{I}\in\Real^{N\times N}$, 
$\vec{s}_c\in\Complex^{N}$ is the $c$th coil-sensitivity map,
%that equals $\vec{S}=\vec{I}$ when $C=1$, 
and 
$\vec{w} \sim \mc{N}(\vec{0}, \vec{I}/\gamma_w)$ is additive white Gaussian noise (AWGN) with precision $\gamma_w$ (i.e., variance $1/\gamma_w$).
In the special case of single-coil MRI, $C=1$ and $\vec{s}_1=\vec{1}$, the all-ones vector.
In this paper, we use the variable-density sampling masks like that shown in \figref{mri_mask}.
In MRI, the pixels-to-measurement ratio, $R\defn N/M$, is known as the ``acceleration rate.''
When $R>1$, one cannot uniquely determine $\vec{x}\true$ from $\vec{y}$ due to the nullspace of $\vec{A}$, and so prior information about $\vec{x}\true$ is needed for its recovery.

To recover images from undersampled MRI measurements, many methods have been proposed.
Some are based on iterative optimization \cite{Lustig:SPM:08,Fessler:SPM:20}, as described in the sequel.
More recently, deep neural networks (DNNs) that directly map MRI measurements $\vec{y}$ to an image $\hvec{x}$ have been proposed, e.g., 
\cite{Jin:TIP:17,Hammernik:MRM:18}.
% Zbontar:18 % fastMRI
Although such DNNs work well, training them requires huge fully-sampled k-space datasets (which may be difficult or impossible to obtain) and changes in the acquisition parameters (e.g., sampling mask) from training to testing can degrade performance \cite{Ahmad:SPM:20}.

In this work, we focus on the ``plug and play'' approach that iteratively calls a DNN for image \emph{denoising}, which brings several advantages.
First, DNN denoisers can be trained using image \emph{patches}, implying the need for relatively few images and no k-space data. 
Second, the denoiser is trained independently of the acquisition parameters, so that it generalizes to any acquisition scenario.
%This approach is known as ``plug-and-play'' (PnP) image recovery \cite{Venkatakrishnan:GSIP:13}. 
%A recent review of plug-and-play methods for MRI is \cite{Ahmad:SPM:20}.
Our approach is based on the generalized expectation consistent (GEC) approximation algorithm from \cite{Fletcher:ISIT:16}, which lives in the family of approximate message passing (AMP) algorithms like \cite{Donoho:PNAS:09,Rangan:TIT:19}.
%Experiments with images from 
%\href{http://mridata.org}{\tt\small http://mridata.org}
%show advantages over existing plug-and-play approaches to MRI image recovery.

% mask

\section{Background}

%%%%%%%%%%%%%%%%%%%%%%%%%%%%%%%%%%%%%%%%%%%%%%%%%%%%%%%%%%%%%%%%%%%%%%%%%%%%%%%%
\subsection{Compressed-sensing-based methods} \label{sec:opt}

The conventional approach to MRI image recovery \cite{Lustig:SPM:08,Fessler:SPM:20} is to pose and solve an optimization problem of the form
\begin{align}
\hvec{x} 
&= \arg\min_{\vec{x}} 
\big\{ g_1(\vec{x}) + g_2(\vec{x}) \big\}
\label{eq:opt},
\end{align}
where $g_1(\vec{x})$ promotes measurement fidelity 
and $g_2(\vec{x})$ is an image-based regularizer.
Typical choices are 
\begin{align}
g_1(\vec{x})
&= \tfrac{\gamma_w}{2}\|\vec{y} - \vec{Ax}\|^2
\label{eq:g1}
\end{align}
for \eqref{y} and $g_2(\vec{x})=\lambda\|\vec{\Psi x}\|_1$ with a suitable transform $\vec{\Psi}$ (e.g., wavelet or total-variation) and carefully chosen $\lambda>0$.
Such $g_2$ encourage sparsity in the transform coefficients $\vec{\Psi x}$.

Many algorithms have been proposed to solve \eqref{opt} with convex $g_1$ and $g_2$ \cite{Fessler:SPM:20}.
For example, the alternating directions method of multipliers (ADMM) \cite{Boyd:FTML:11} iterates
\begin{subequations}
\label{eq:admm}
\begin{align}
\vec{x}_1 &\gets \prox_{\gamma^{-1}g_1}(\vec{x}_2-\vec{u})
\qquad
\label{eq:admm_loss}\\
\vec{x}_2 &\gets \prox_{\gamma^{-1}g_2}(\vec{x}_1+\vec{u})
\label{eq:admm_prox}\\
\vec{u} &\gets \vec{u} + \left(\vec{x}_1 - \vec{x}_2\right) 
\label{eq:admm_dual},
\end{align}
\end{subequations}
where the proximal operator is defined as
\begin{align}
\prox_{\rho}(\vec{r}) 
\defn \arg\min_{\vec{x}} \big\{ \rho(\vec{x}) + \tfrac{1}{2}\|\vec{x}-\vec{r}\|^2 \big\}
\label{eq:prox}.
\end{align}
In \eqref{admm}, $\gamma>0$ is a tunable stepsize that affects the speed of ADMM's convergence but not its fixed point.

%%%%%%%%%%%%%%%%%%%%%%%%%%%%%%%%%%%%%%%%%%%%%%%%%%%%%%%%%%%%%%%%%%%%%%%%%%%%%%%%
\subsection{Plug-and-play methods} \label{sec:pnp}

The prox operator \eqref{prox} can interpreted as a denoiser, in particular, the maximum a posteriori (MAP) estimator of $\vec{x}\true$ with prior $p(\vec{x}\true)\propto e^{-g_2(\vec{x}\true)}$ from an observation $\vec{r}=\vec{x}_0+\vec{e}$ with $\gamma$-precision AWGN $\vec{e}$.
Leveraging this fact, Bouman et al.\ \cite{Venkatakrishnan:GSIP:13} proposed to replace ADMM line~\eqref{admm_prox} with a call to a high-performance image denoiser $\vec{f}_2(\vec{x})$ like BM3D \cite{Dabov:TIP:07} 
or DnCNN~\cite{Zhang:TIP:17}, giving rise to ``plug-and-play'' (PnP) ADMM. 
PnP extensions of other algorithms, such as primal-dual splitting (PDS) \cite{Ono:SPL:17} and proximal gradient (PG) \cite{Kamilov:SPL:17}, 
have also been proposed.
As shown in the recent overview paper, PnP methods have been shown to significantly outperform compressed-sensing-based approaches in MRI \cite{Ahmad:SPM:20}.
Note, however, that when \eqref{admm_prox} is replaced with a denoising step of the form ``$\vec{x}_2\gets\vec{f}_2(\vec{x}_1+\vec{u})$,'' the stepsize $\gamma$ \emph{does} affect the fixed-point \cite{Ahmad:SPM:20} and thus must be tuned.

Although PnP algorithms work well in MRI, there is room for improvement.  
For example, while image denoisers are typically designed/trained to remove the effects of AWGN, PnP algorithms do not provide the denoiser with an AWGN-corrupted input at each iteration.
Rather, the denoiser's input error has iteration-dependent statistics that are difficult to analyze or predict.

%PnP-FISTA~\cite{Kamilov:SPL:17}, RED~\cite{Romano:JIS:17, Reehorst:TCI:19}, PnP-PDS~\cite{Ono:SPL:17} are examples of some other popular PnP algorithms. 

%%%%%%%%%%%%%%%%%%%%%%%%%%%%%%%%%%%%%%%%%%%%%%%%%%%%%%%%%%%%%%%%%%%%%%%%%%%%%%%%
\subsection{Approximate message passing}

In \eqref{opt}, if we interpret $g_1(\vec{x})$ as a log-likelihood and $g_2(\vec{x})$ as a log prior, then $\hvec{x}$ can be interpreted as the MAP estimate of $\vec{x}$ from $\vec{y}$.
However, because image recovery results are often judged by mean-squared error (MSE), one may be more interested in the minimum MSE (MMSE) estimate of $\vec{x}$ from $\vec{y}$.
Interestingly, both MMSE and MAP estimation are facilitated by approximate message passing (AMP) methods like \cite{Donoho:PNAS:09,Rangan:TIT:19}.
%belief-propagation-based methods \cite{Wainwright:FTML:08}  

For example, the AMP algorithm from \cite{Donoho:PNAS:09} iterates 
\begin{subequations}
\label{eq:amp}
\begin{eqnarray}
\vec{v}^\iters 
&=& \beta\cdot\big( \vec{y}-\vec{Ax}^\iter 
        + \tfrac{1}{M}\vec{v}^\iter \tr\{\nabla\vec{f}_2^\iter(\vec{x}^\itero\!+\!\vec{A}\herm\vec{v}^\iter)\} \big)
\qquad
\label{eq:amp_onsager}\\
\tau^\iters
&=& \tfrac{1}{M}\|\vec{v}^\iters\|^2
\label{eq:amp_noisevar}\\
%\vec{r}^\iters 
%&=& \vec{x}^\iter + \vec{A}\herm\vec{v}^\iters \\
\vec{x}^\iters
%&=\vec{f}_2^\iters(\vec{r}^\iters)
&=&\vec{f}_2^\iters(\vec{x}^\iter + \vec{A}\herm\vec{v}^\iters)
\end{eqnarray}
\end{subequations}
over $t=0,1,2,\dots$, starting from $\vec{v}^0=\vec{0}=\vec{x}^0$, where 
$\vec{f}_2^\iter(\cdot)$ is a Lipschitz denoising function,
$\tr\{\nabla\vec{f}_2^\iter(\vec{r})\}$ is the trace of the Jacobian of $\vec{f}_2^\iter$ at $\vec{r}$,
and $\beta=N/\|\vec{A}\|_F^2$.
When configured for MAP estimation, AMP uses the MAP denoiser $\vec{f}_2^\iter(\vec{r})=\prox_{\tau^\iter g_2}(\vec{r})$.
When configured for MMSE estimation, AMP instead uses the MMSE denoiser $\vec{f}_2^\iter(\vec{r})=\E\{\vec{x}\,|\,\vec{r}\}$ for $\vec{r}=\vec{x}+\vec{w}$ with $\vec{w}\sim\mc{N}(\vec{0},\tau^t\vec{I})$.

Importantly, when the forward operator $\vec{A}$ is large and i.i.d.\ sub-Gaussian, and when $\vec{f}_2^\iter$ is Lipschitz, the macroscopic behavior of AMP is rigorously characterized by a scalar state-evolution \cite{Berthier:II:19,Bayati:TIT:11}.
When $\vec{f}_2^t$ is the MMSE denoiser and the state-evolution has a unique fixed point, AMP provably converges to the MMSE-optimal estimate $\hvec{x}\MMSE$ \cite{Berthier:II:19,Bayati:TIT:11}.
For images, the MMSE denoiser can be approximated by BM3D or a DNN, as proposed in \cite{Metzler:ICIP:15}, leading to ``denoising-AMP'' (D-AMP).
There, the trace-Jacobian in \eqref{amp_onsager} is approximated using the Monte-Carlo approach \cite{Ramani:TIP:08}
\begin{align}
\tr\{\nabla\vec{f}_2^\iter(\vec{r})\}
\approx \delta^{-1}\vec{q}\herm\big[\vec{f}_2^\iter(\vec{r}+\delta\vec{q})-\vec{f}_2^\iter(\vec{r)}\big]
\label{eq:trJfapprox} ,
\end{align}
with random $\vec{q}\sim\mc{N}(\vec{0},\vec{I})$ and small $\delta>0$.

More recently, the vector AMP (VAMP) algorithm \cite{Rangan:TIT:19} was proposed, with similar properties as AMP (e.g., rigorous state evolution and provable MMSE estimation) but applicability to a wider class of random matrices: right orthogonally invariant (ROI) ones.
%An ROI matrix has a singular value decomposition of the form $\vec{USV}\herm$, where $\vec{U}$ is unitary, $\vec{S}$ is diagonal, and $\vec{V}$ is Haar (i.e., uniformly distributed over the set of unitary matrices). 
Inspired by D-AMP, a denoising VAMP (D-VAMP) was proposed in \cite{Schniter:BASP:17} and analyzed in \cite{Fletcher:NIPS:18}.

%%%%%%%%%%%%%%%%%%%%%%%%%%%%%%%%%%%%%%%%%%%%%%%%%%%%%%%%%%%%%%%%%%%%%%%%%%%%%%%%
\subsection{AMP for MRI}

Neither AMP nor VAMP works as intended in MRI because $\vec{A}$ in \eqref{y} lacks sufficient randomness.
In fact, these algorithms tend to diverge in MRI if applied without modification.
%when $\vec{A}$ and $\vec{x}\true$ both have a Fourier structure.

The failure of AMP and VAMP can be understood from their error recursions. For AMP, the error recursion is \cite{Schniter:TSP:20}
\begin{eqnarray}
\vec{e}^\iters 
&=& (\vec{I}\!-\!\vec{A}\tran\vec{A})\vec{\epsilon}^\iter 
   + \vec{A}\tran(\vec{w}\!+\!\tfrac{1}{M}\vec{v}^\iter
     \tr\{\nabla\vec{f}_2^\iter(\vec{x}\true\!+\!\vec{e}^\iter)\}) \quad 
     \label{eq:err_iter1}\\
\vec{\epsilon}^\iters
&=& \vec{f}_2^\iters(\vec{x}\true+\vec{e}^\iters)-\vec{x}\true
     \label{eq:err_iter2}.
\end{eqnarray}
It is important to keep in mind that images $\vec{x}\true$ have much more energy at low Fourier frequencies than at high ones. 
The same tends to be true of the output error $\vec{\epsilon}^\iter$ of an image denoiser.
Even so, if $\vec{A}\in\Real^{M\times N}$ was large and i.i.d.\ (with zero mean and elementwise variance $\frac{1}{M}$), then the $\vec{I}-\vec{A}\tran\vec{A}$ term in \eqref{err_iter1} would randomize $\vec{\epsilon}^\iter$ such that the denoiser input error vector $\vec{e}^\iters$ looks like AWGN.
%The AWGN property of $\vec{e}^\iters$ is then leveraged in the next iteration.
%VAMP shows a similar behavior when $\vec{A}$ is ROI and high-dimensional.
In MRI, however, both $\vec{A}$ and $\vec{\epsilon}^\iter$ have Fourier structure, this randomization does not happen, and AMP behaves unpredictably.
A similar behavior plagues VAMP.

Several MRI-specific variations of AMP and VAMP have been proposed to counter these deficiencies.
For example, \cite{Eksioglu:JIS:18} proposed D-AMP with a very small $\beta$, which helps the algorithm converge, but at the cost of degrading its fixed points.
%A refinement based on a matrix-valued $\beta$ was proposed in \cite{Sarkar:Diss:20}.
\cite{Sarkar:ICASSP:21} proposed a damped D-VAMP that, combined with a novel initialization, showed improved performance and runtime over PnP-ADMM for MRI.

Several other VAMP-based algorithms for MRI have been designed to recover the wavelet coefficients of the image rather than the image itself.
The motivation is that, in this case, $\vec{A}$ is a Fourier-Wavelet matrix, which is approximately block diagonal \cite{Adcock:FMS:17}, where the blocks correspond to the wavelet subbands.
With an appropriate modification of VAMP, the subband error vectors can be made to behave more like AWGN, albeit with different variances.
The first incarnation of this idea appeared in \cite{Schniter:BASP:17b}, where a fixed bandwise normalization procedure was used. 
Later, for single-coil MRI with variable-density sampling masks, a ``variable density AMP'' (VDAMP) algorithm with band-specific adaptive wavelet thresholding was proposed in \cite{Millard:20}, which was able to successfully predict the noise variance in each subband at each iteration.
More recently, the D-VDAMP algorithm \cite{Metzler:ICASSP:21} extended VDAMP to DNN denoising in each subband.

Although D-VDAMP is the state-of-the-art AMP algorithm for MRI, it is based on a non-standard modification of VAMP with degraded fixed points, which makes early stopping critical for good performance.
Also, it is not clear how to extend D-VDAMP to multi-coil MRI.
These issues motivate our approach, which is described next.

%%%%%%%%%%%%%%%%%%%%%%%%%%%%%%%%%%%%%%%%%%%%%%%%%%%%%%%%%%%%%%%%%%%%%%%%%%%%%%%%
\section{Proposed Approach}

\subsection{Denoising GEC} \label{sec:dgec}

Our approach uses the GEC framework from \cite{Fletcher:ISIT:16}, which is summarized in \algref{gec}.
When solving a convex optimization problem of the form \eqref{opt}, the functions $\vec{f}_i$ in \algref{gec} take the form
\begin{align}
\vec{f}_i(\vec{r},\vec{\gamma}) 
&= \gprox_{g_i,\vec{\gamma}}(\vec{r}), ~~i=1,2,
\end{align}
for the generalized proximal operator
\begin{align}
\gprox_{\rho,\vec{\gamma}}(\vec{r}) 
&\defn \arg\min_{\vec{x}} \big\{ \rho(\vec{x}) + \tfrac{1}{2}\|\vec{x}-\vec{r}\|^2_{\vec{\gamma}}\big\}
\label{eq:gprox} ,
\end{align}
where $\|\vec{q}\|_{\vec{\gamma}} \defn \sqrt{\vec{q}\herm\Diag(\vec{\gamma})\vec{q}}$
and $\Diag(\cdot)$ creates a diagonal matrix from its vector argument.
Note that if $\vec{\gamma}=\gamma\vec{1}$, then $\gprox_{\rho,\vec{\gamma}}=\prox_{\gamma^{-1}\rho}$.
Furthermore, if the $\vec{\gamma}_i$ vectors were held fixed over the iterations and took the form $\vec{\gamma}_i=\gamma_i\vec{1}$, then \algref{gec} reduces to a variant of ADMM \eqref{admm} with two dual updates: \eqref{admm_dual} and a similar step between \eqref{admm_loss} and \eqref{admm_prox}.
So, GEC can be interpreted as an ADMM-like algorithm with two adaptive vector-valued stepsizes, $\vec{\gamma}_1$ and $\vec{\gamma}_2$.

\begin{algorithm}
\caption{Generalized EC (GEC)}
\label{alg:gec}
\begin{algorithmic}[1]  
\Require{$\vec{f}_1(\cdot,\cdot), \vec{f}_2(\cdot,\cdot),\text{ and }\gdiag(\cdot)$.}
\State{Select initial $\vec{r}_1,\vec{\gamma}_1$}
\Repeat
    \State{// Measurement fidelity}
    \State{$\hvec{x}_1 \gets \vec{f}_1(\vec{r}_1,\vec{\gamma}_1)$} \label{line:x1}
    \State{$\vec{\eta}_1 \gets 
        \Diag( \gdiag( \nabla\vec{f}_1(\vec{r}_1,\vec{\gamma}_1) ))^{-1} 
        \vec{\gamma}_1$} \label{line:eta1}
    \State{$\vec{\gamma}_2 \gets \vec{\eta}_1 - \vec{\gamma}_1$}  \label{line:gam2}
    \State{$\vec{r}_2 \gets \Diag(\vec{\gamma}_2)^{-1}(\Diag(\vec{\eta}_1)\hvec{x}_1 - \Diag(\vec{\gamma}_1)\vec{r}_1)$}
        \label{line:r2}
    \State{// Denoising}
    \State{$\hvec{x}_2 \gets \vec{f}_2(\vec{r}_2,\vec{\gamma}_2)$} \label{line:x2}
    \State{$\vec{\eta}_2 \gets 
        \Diag( \gdiag( \nabla\vec{f}_2(\vec{r}_2,\vec{\gamma}_2) ))^{-1} 
        \vec{\gamma}_2$} \label{line:eta2}
    \State{$\vec{\gamma}_1 \gets \vec{\eta}_2 - \vec{\gamma}_2$}  \label{line:gam1}
    \State{$\vec{r}_1 \gets \Diag(\vec{\gamma}_1)^{-1}(\Diag(\vec{\eta}_2)\hvec{x}_2 - \Diag(\vec{\gamma}_2)\vec{r}_2)$}
        \label{line:r1}
\Until{Terminated}
\end{algorithmic}
\end{algorithm}

In lines~\ref{line:eta1} and \ref{line:eta2}, GEC averages the diagonal of the Jacobian separately over $L$ coefficient subsets using $\gdiag\!:\Real^{N\times N}\!\rightarrow\!\Real^N$: 
\begin{align}
\gdiag(\vec{Q}) \defn [d_1\vec{1}_{N_1}\tran,\dots,d_L\vec{1}_{N_L}\tran]\tran,
~d_\ell = \frac{\tr\{\vec{Q}_{\ell\ell}\}}{N_\ell}
\label{eq:gdiag} .
\end{align}
In \eqref{gdiag}, $N_\ell$ is the size of the $\ell$th subset
%(so that $\sum_{\ell=1}^L N_\ell=N$), 
%$\vec{1}_{N_\ell}$ is the ones vector of dimension $N_\ell$, 
and $\vec{Q}_{\ell\ell}\in\Real^{N_\ell\times N_\ell}$ is the $\ell$th diagonal subblock of the matrix input $\vec{Q}$.
When $L\!=\!1$, GEC reduces to VAMP.

We focus on the quadratic loss \eqref{g1}, which yields
\begin{align}
\vec{f}_1(\vec{r},\vec{\gamma}) 
&= \big(\gamma_w\vec{A}\herm\vec{A}+\Diag(\vec{\gamma})\big)^{-1}\big(\gamma_w\vec{A}\herm\vec{y}+\Diag(\vec{\gamma})\vec{r}\big)
\label{eq:f1} .
\end{align}
For $\vec{f}_2$, we propose to ``plug in'' a DNN denoiser. 
For both $\vec{f}_1$ and $\vec{f}_2$, we approximate the $\tr\{\vec{Q}_{\ell\ell}\}$ term in \eqref{gdiag} using 
\begin{eqnarray}
\tr\{\vec{Q}_{\ell\ell}\}
&\approx& \delta^{-1}\vec{q}_\ell\herm\big[\vec{f}_i(\vec{r}+\delta\vec{q}_\ell,\vec{\gamma})-\vec{f}_i(\vec{r},\vec{\gamma})\big]
\label{eq:trJfapprox_ll} ,
\end{eqnarray}
where the $\ell$th coefficient subset in $\vec{q}_\ell$ is i.i.d.\ unit-variance Gaussian and the others are zero.
Inspired by D-AMP and D-VAMP, we call this approach ``denoising GEC'' (D-GEC).

\subsection{D-GEC for Image Recovery} \label{sec:wdgec}

Like \cite{Schniter:BASP:17b,Millard:20,Metzler:ICASSP:21}, we recover the wavelet coefficients $\vec{c}\true$ rather than the image pixels $\vec{x}\true$.
For orthogonal wavelet transform $\vec{\Psi}$, we have
$\vec{c}\true = \vec{\Psi x}\true$ 
and 
$\vec{x}\true = \vec{\Psi}\tran\vec{c}\true$,
so that we can rewrite \eqref{y} as
\begin{align}
\vec{y} = \vec{B}\vec{c}\true + \vec{w}
\text{~~with~~}
\vec{B} \defn \vec{A\Psi}\tran 
\label{eq:y2} .
\end{align}
To apply D-GEC to $\vec{c}\true$-recovery, we choose $\vec{f}_1$ as in \eqref{f1}, but with $\vec{B}$ in place of $\vec{A}$, and for the diagonalization subsets we choose the $L=3D\!+\!1$ subbands of a depth-$D$ 2D wavelet transform.
As in \cite{Metzler:ICASSP:21}, we perform denoising in the wavelet domain using a denoiser $\vec{f}_2$ that can exploit knowledge of the noise variance in each wavelet subband, as provided by the precision vector $\vec{\gamma}_2$.

\section{Numerical Experiments}

We now compare the proposed D-GEC algorithm to the existing D-VDAMP \cite{Metzler:ICASSP:21} and PnP-PDS \cite{Ono:SPL:17} algorithms.
Based on the extensive experiments in \cite{Metzler:ICASSP:21}, D-VDAMP is state-of-the-art among PnP algorithms. 
PnP-PDS is a useful baseline, since it has
the same fixed points as PnP-ADMM and PnP-PG.

\noindent\textbf{Denoisers:}
For a fair comparison to D-VDAMP \cite{Metzler:ICASSP:21}, we use the DNN denoiser proposed in \cite{Metzler:ICASSP:21}, which is modification of DnCNN \cite{Zhang:TIP:17} that accepts the noise standard deviation (SD) in each wavelet subband. 
The denoiser was trained using noise that was white in each subband but with SD that varies across subbands.  
In particular, 5 copies of the denoiser were trained using subband noise SDs uniformly distributed in the ranges 0-10, 10-20, 20-50, 50-120, and 120-500, respectively.
(Pixel values ranged from 0-255.) 
The DNNs were trained using 
%61\,882 different patches formed by cropping, scaling, flipping, and rotating 
patches from
70 MRI images of the Stanford 2D FSE dataset available at \href{http://mridata.org}{\small \tt http://mridata.org}.
For PnP-PDS, we used a standard DnCNN denoiser trained on the same data with white noise of SD uniformly distributed in 20-50, as in \cite{Metzler:ICASSP:21}.
Because we used real-valued images, the denoisers use only the real part of the input and generate a real-valued output.

\begin{figure}[t]
	\centering
	\newcommand{\wid}{0.09\textwidth}
	\includegraphics[width=\wid]{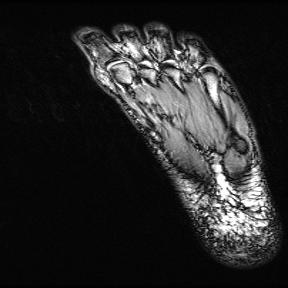}
	\includegraphics[width=\wid]{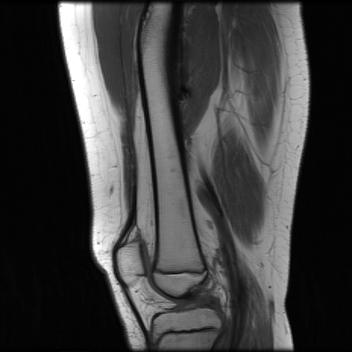}
	\includegraphics[width=\wid]{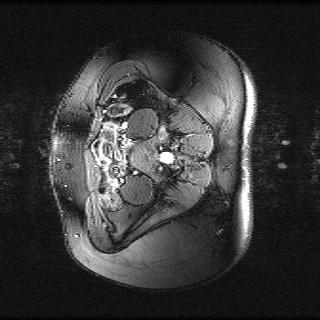}
	\includegraphics[width=\wid]{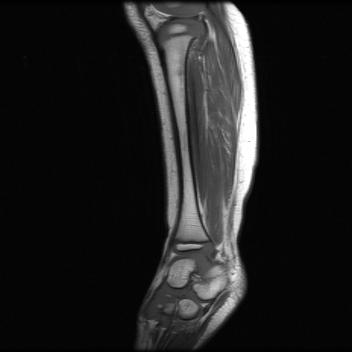}
	\includegraphics[width=\wid]{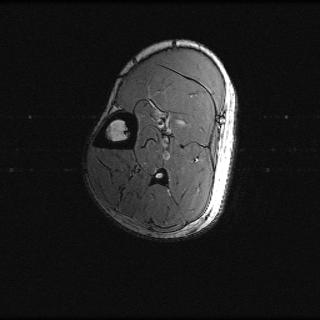}~\mbox{}\\
	\includegraphics[width=\wid]{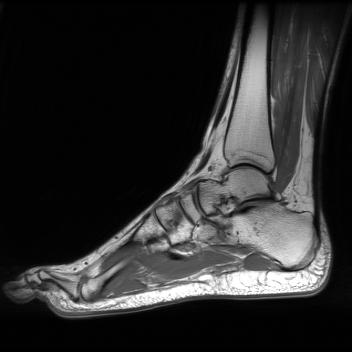}
	\includegraphics[width=\wid]{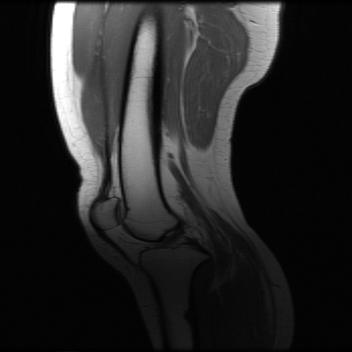}
	\includegraphics[width=\wid]{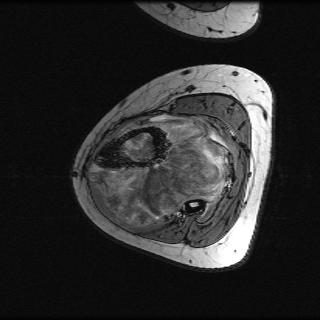}
	\includegraphics[width=\wid]{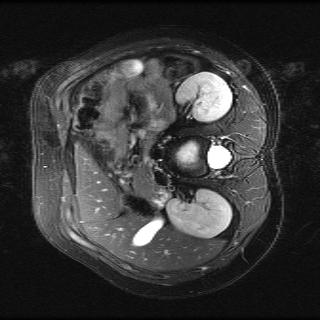}
	\includegraphics[width=\wid]{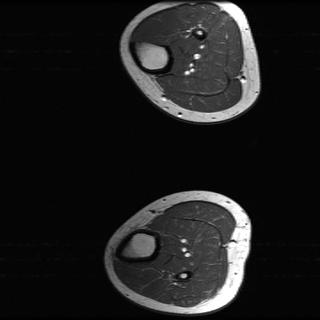}
	\vspace{-2mm}
	\caption{Test images from \href{http://mridata.org}{\small \tt http://mridata.org}.}
	\label{fig:test_images}
	\vspace{-4mm}
\end{figure}

\noindent\textbf{Test data:}
For evaluation, we used the ten 352$\times$352 MRI images in \figref{test_images}, which were not in the training dataset.
The measurements $\vec{y}$ were constructed using \eqref{y} with complex AWGN $\vec{w}$ whose variance was adjusted to give a pre-masking SNR of 40 dB.
For the multicoil experiments, we used coil sensitivities $\vec{s}_c$ simulated using the Biot-Savart law, while in the single-coil case, we used $\vec{s}_1=\vec{1}$. 

% NMSE evolution 
\begin{figure*}[t!]
	\centering
	%\includegraphics[width=\columnwidth]{figures/single_coil/SC_VD_R4_PreMSNR40_NMSE_Evolution_D_GEC.png}
	%\vspace{-7mm}
	%\includegraphics[width=\columnwidth,trim=50 410 70 30,clip]{figures/single_coil/SC_VD_R4_PreMSNR40_MSEandNMSE_Evolution_D_GEC.png}
	%\includegraphics[width=\columnwidth,trim=0 10 0 5,clip]{figures/single_coil/SC_VD_R4_PreMSNR40_SD_Evolution_D_GEC.png}
%	\includegraphics[width=0.8\textwidth,trim=0 10 0 5,clip]{figures/single_coil/SC_VD_R4_PreMSNR40_SD_Evolution_D_GEC.png}
	\includegraphics[width=0.9\textwidth,trim=0 10 0 5,clip]{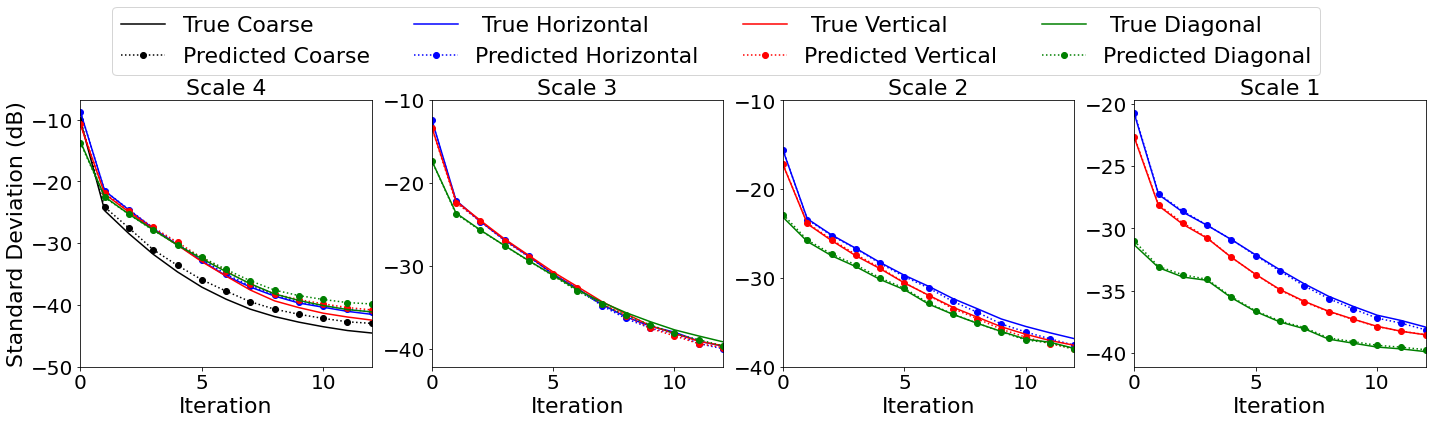}
	\vspace{-2mm}
	\caption{SD of D-GEC's denoiser input versus iteration.}
	\label{fig:mse_evo}
	\vspace{-3mm}
\end{figure*}

% image plots
\begin{figure*}[t!]
	\newcommand{\wid}{0.21\textwidth}
	\centering
	%\boxed{
		\includegraphics[width=0.216\textwidth,trim=10 2 5 10,clip]{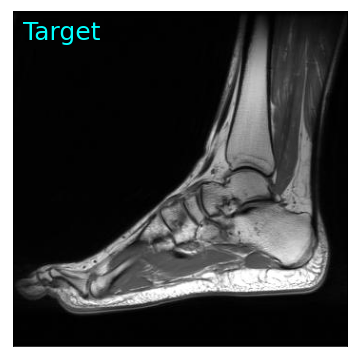}
		%}
	%\boxed{
		\includegraphics[width=\wid,trim=15 10 25 10,clip]{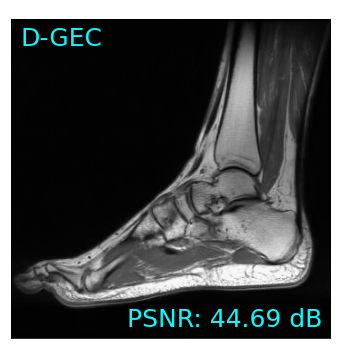}
		%}
	\includegraphics[width=\wid,trim=15 10 25 10,clip]{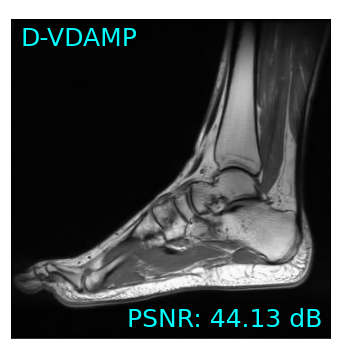}
	\includegraphics[width=\wid,trim=15 10 25 10,clip]{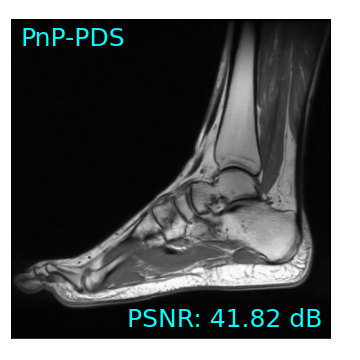}\\
	\includegraphics[width=0.215\textwidth,trim=0 -290 0 0]{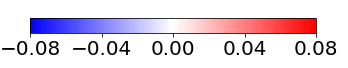}
	\includegraphics[width=\wid,trim=15 14 25 14,clip]{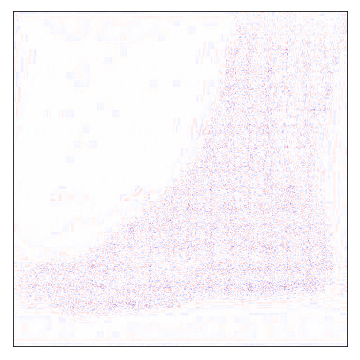}
	\includegraphics[width=\wid,trim=15 14 25 14,clip]{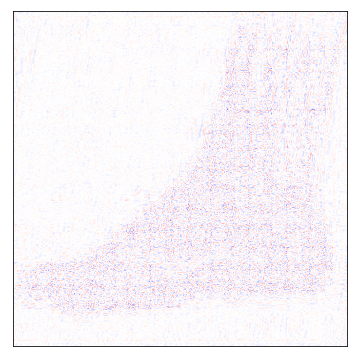}
	\includegraphics[width=\wid,trim=15 14 25 14,clip]{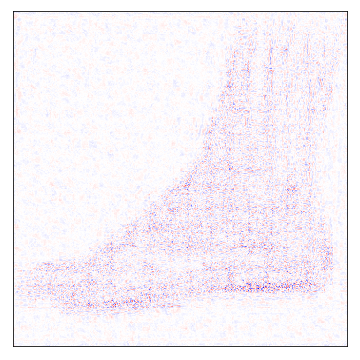}
	\vspace{-2mm}
	\caption{Example single-coil recoveries and error maps.}
	\label{fig:example}
	\vspace{-4mm}
\end{figure*}

% wavelet coefficient plots
\begin{figure}[ht]
\newcommand{\wid}{0.32\columnwidth}
\centering
  \hspace{0.09\columnwidth}
  \quad$\vec{c}\true$\quad
  \hfill
  \quad$\vec{r}_2$\quad
  \hfill
  $|\vec{r}_2-\vec{c}\true|$
  \hspace{0.09\columnwidth}\mbox{}\\[-0.5mm]
\includegraphics[width=\wid]{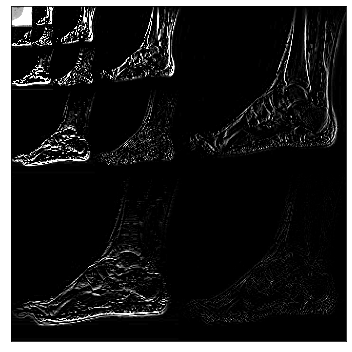}
\includegraphics[width=\wid]{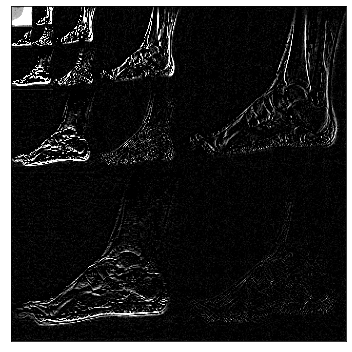}
\includegraphics[width=\wid]{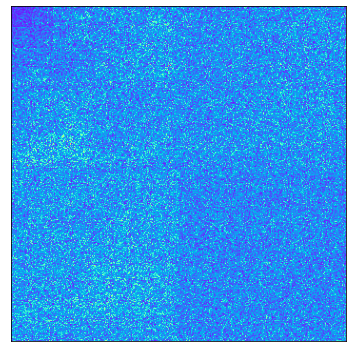}
\vspace{-2mm}
\caption{True coefficients, denoiser input, error (iteration 10).}
\label{fig:wavelet_coefs}
\vspace{-2mm}
\end{figure}

% QQ plots
\begin{figure}[ht]
\centering
\hspace{0.01\columnwidth}
\quad Horizontal, Scale 3 \quad
\hfill \hspace{-0.07\columnwidth}
\quad Vertical, Scale 2 \quad
\hfill 
Diagonal, Scale 1 
\hspace{0.00\columnwidth}\mbox{}\\ %[-0.5mm]
\includegraphics[width=\columnwidth,trim=6 8 16 25,clip]{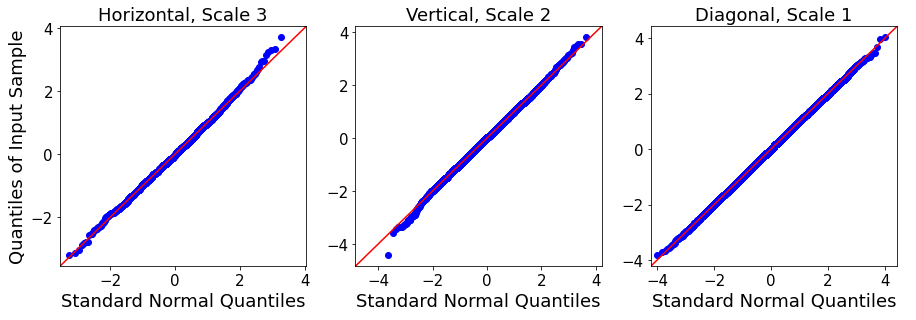}
\vspace{-5mm}
\caption{Wavelet coefficient QQ-plots showing Gaussianity.}
\label{fig:qq_plots}
\end{figure}

\noindent\textbf{Algorithm parameters:}
For D-GEC and D-VDAMP, we used a 2D Haar wavelet transform with $D\!=\!4$ levels, giving $L\!=\!13$ wavelet subbands. 
D-GEC used the auto-tuning scheme from \cite{Fletcher:NIPS:17} and the damping scheme from \cite{Sarkar:ICASSP:21} with parameter $0.4$.
D-VDAMP code was obtained from the authors and run under default settings, which are detailed in \cite{Metzler:ICASSP:21}.
PnP-PDS was run for 200 iterations using the stepsize that maximized PSNR on the training set.

% PSNR and SSIM results
\begin{table}[t!]
	\centering
	\resizebox{\columnwidth}{!}{
		\begin{tabular}{@{}c|cc|cc|cc|cc|}
			& \multicolumn{4}{c|}{$C=1$ coil} & \multicolumn{4}{c|}{$C=4$ coils} \\
			& \multicolumn{2}{c|}{$R=4$} & \multicolumn{2}{c|}{$R=8$} 
			& \multicolumn{2}{c|}{$R=4$} & \multicolumn{2}{c|}{$R=8$} \\
			method  & PSNR    & SSIM    & PSNR & SSIM & PSNR    & SSIM    & PSNR & SSIM\\\hline
			PnP-PDS & 40.66   & 0.968   &  37.38     &  \bf 0.951    & 34.71 & 0.935 &  33.09 & 0.917     \\ 
			D-VDAMP & 42.36   & 0.972   &  35.92    &   0.918   & n/a     & n/a     & n/a  & n/a  \\
			D-GEC   &\bf 42.97&\bf 0.977&   \bf 37.65   &   0.946   &\bf 45.18 &\bf 0.993&  \bf 41.13   &   \bf 0.982   \\
	\end{tabular}}
	\vspace{-2mm}
	\caption{Recovery results averaged over the 10 test images.}
	\label{tab:results}
	\vspace{-3mm}
\end{table}

\noindent\textbf{Single-coil results:}
\tabref{results} shows that D-GEC outperformed D-VDAMP in all single-coil experiments and outperformed PnP-PDS in all but SSIM at $R\!=\!8$.
\Figref{wavelet_coefs} shows an example of the wavelet coefficients input to D-GEC's denoiser at the 10th iteration, and their error relative to the true coefficients.
\Figref{mse_evo} shows the evolution of the standard deviation at the input to D-GEC's denoiser in each subband; there is a good agreement between true and predicted values.
\Figref{qq_plots} suggests that the subband errors are Gaussian.
\Figref{example} shows image recoveries and error maps for one test image at $R=4$.

\noindent\textbf{Multi-coil results:}
\tabref{results} shows D-GEC significantly outperforming PnP-PDS in PSNR and SSIM in the 4-coil case.
D-VDAMP does not support multi-coil recovery and thus is not shown.

\section{Conclusion}

We designed a GEC-based plug-and-play algorithm for MRI that renders the subband errors white and Gaussian with predictable variance, and used it with a denoiser trained to handle subband errors that are white and Gaussian with known variance.
Experiments show good performance relative to previous approaches in single- and multi-coil settings.

%%%%%%%%%%%%%%%%%%%%%%%%%%%%%%%%%%%%%%%%%%%%%%%%%%%%%%%%%%%%%%%%%%%%%%%%%%%%%%%%
% control near the beginning of the file
\iftoggle{extra}{
\clearpage

\textr{THIS PAGE CONTAINS EXTRA RESULTS, AND WILL BE REMOVED BEFORE SUBMISSION!}
 \begin{table}[ht]
  \centering
  \caption{PSNR and SSIM comparison averaged over the $10$ test images for two different pre-mask measured SNRs with acceleration R = $4$ and single coil}
  \begin{tabular}{ | c|| c| c|| c| c|}
  \hline
  \multicolumn{1}{|c||}{} &\multicolumn{2}{c||}{SNR: $30$ dB}  &\multicolumn{2}{c||}{SNR: $40$ dB} \\
 \cline{2-5}
  \multicolumn{1}{|c||}{method} & \multicolumn{1}{c|}{PSNR(dB)} & \multicolumn{1}{c||}{SSIM} & \multicolumn{1}{c|}{PSNR(dB)} & \multicolumn{1}{c||}{SSIM} \\
  \hline
 \hline
 PnP-PDS & $39.40$ & $0.64$ & $40.66$ &   $0.968$\\
 \hline
 D-VDAMP & $40.36$ & $0.965$  & $42.36$ &  $0.972$    \\
 \hline
 D-GEC & $\mathbf{40.89}$ &   $\mathbf{0.971}$ &  $\mathbf{42.97}$ &   $\mathbf{0.977}$  \\
 \hline
 \end{tabular}
 \label{tab:results_1}
 \end{table}

 \begin{table}[ht]
  \centering
  \caption{PSNR and SSIM comparison averaged over the $10$ test images for two different accelerations with pre-mask measured SNR of $40$ dB and single coil}
  \begin{tabular}{ | c|| c| c|| c| c|}
  \hline
  \multicolumn{1}{|c||}{} &\multicolumn{2}{c||}{$R\!=\!8$}  &\multicolumn{2}{c||}{$R\!=\!6$} \\
 \cline{2-5}
  \multicolumn{1}{|c||}{method} & \multicolumn{1}{c|}{PSNR(dB)} & \multicolumn{1}{c||}{SSIM} & \multicolumn{1}{c|}{PSNR(dB)} & \multicolumn{1}{c||}{SSIM} \\
  \hline
 \hline
 PnP-PDS & $37.38$ & $\mathbf{0.951}$ & $38.81$ &  $0.959$\\
 \hline
 D-VDAMP & $35.92$ & $0.918$  & $38.58$ &  $0.943$    \\
 \hline
 D-GEC & $\mathbf{37.65}$ &   $0.946$ & $\mathbf{40.17}$ &   $\mathbf{0.965}$  \\
 \hline
 \end{tabular}
 \label{tab:results_2}
 \end{table}

 \begin{table}[ht]
  \centering
  \caption{PSNR and SSIM comparison averaged over the $10$ test images for two different accelerations with pre-mask measured SNR of $40$ dB and $C = 4$ coils}
  \begin{tabular}{ | c|| c| c|| c| c|}
  \hline
  \multicolumn{1}{|c||}{} &\multicolumn{2}{c||}{$R\!=\!8$}  &\multicolumn{2}{c||}{$R\!=\!4$} \\
 \cline{2-5}
  \multicolumn{1}{|c||}{method} & \multicolumn{1}{c|}{PSNR (dB)} & \multicolumn{1}{c||}{SSIM} & \multicolumn{1}{c|}{PSNR (dB)} & \multicolumn{1}{c||}{SSIM} \\
  \hline
 \hline
 PnP-PDS & $32.91$ & $0.917$ & $34.65$ &   $0.935$\\
 \hline
 D-GEC & $\mathbf{39.20}$ &   $\mathbf{0.974}$  & $\mathbf{44.06}$ &   $\mathbf{0.990}$  \\
 \hline
 \end{tabular}
 \label{tab:results_3}
 \end{table}
}%\iftoggle

%%%%%%%%%%%%%%%%%%%%%%%%%%%%%%%%%%%%%%%%%%%%%%%%%%%%%%%%%%%%%%%%%%%%%%%%%%%%%%%%
\clearpage
\bibliographystyle{IEEEtran}

\bibliography{macros_abbrev,phase,books,misc,comm,multicarrier,sparse,machine}

%\bibliography{\myreferences/macros_abbrev,\myreferences/phase,\myreferences/books,\myreferences/misc,\myreferences/comm,\myreferences/multicarrier,\myreferences/sparse,\myreferences/machine}

%\bibliography{/Users/sauravkumaraswamishastri/OneDrive/git/group_git/bib/macros_abbrev,/Users/sauravkumaraswamishastri/OneDrive/git/group_git/bib/phase,/Users/sauravkumaraswamishastri/OneDrive/git/group_git/bib/books,/Users/sauravkumaraswamishastri/OneDrive/git/group_git/bib/misc,/Users/sauravkumaraswamishastri/OneDrive/git/group_git/bib/comm,/Users/sauravkumaraswamishastri/OneDrive/git/group_git/bib/multicarrier,/Users/sauravkumaraswamishastri/OneDrive/git/group_git/bib/sparse,/Users/sauravkumaraswamishastri/OneDrive/git/group_git/bib/machine} 

\end{document}